\documentclass[twocolumn]{aastex631}

\usepackage{xcolor}

\begin{document}

\title{SMA detection of an extreme millimeter flare from the young class~III star HD~283572}

\correspondingauthor{Joshua Bennett Lovell}; \email{joshualovellastro@gmail.com}

\author[0000-0002-4248-5443]{Joshua Bennett Lovell}
\affiliation{Center for Astrophysics, Harvard \& Smithsonian, 60 Garden Street, Cambridge, MA 02138-1516, USA}

\author[0000-0002-3490-146X]{Garrett K. Keating}
\affiliation{Center for Astrophysics, Harvard \& Smithsonian, 60 Garden Street, Cambridge, MA 02138-1516, USA}

\author[0000-0003-1526-7587]{David J. Wilner}
\affiliation{Center for Astrophysics, Harvard \& Smithsonian, 60 Garden Street, Cambridge, MA 02138-1516, USA}

\author[0000-0003-2253-2270]{Sean M. Andrews}
\affiliation{Center for Astrophysics, Harvard \& Smithsonian, 60 Garden Street, Cambridge, MA 02138-1516, USA}

\author[0000-0001-7891-8143]{Meredith MacGregor}
\affiliation{Department of Physics and Astronomy, Johns Hopkins University, Baltimore, MD 21218, USA}

\author[0009-0003-3708-0518]{Ramisa Akther Rahman}
\affiliation{Center for Astrophysics, Harvard \& Smithsonian, 60 Garden Street, Cambridge, MA 02138-1516, USA}
\affiliation{William \& Mary College, 15 Richmond Rd, Williamsburg, VA, USA}

\author[0000-0002-1407-7944]{Ramprasad Rao}
\affiliation{Center for Astrophysics, Harvard \& Smithsonian, 60 Garden Street, Cambridge, MA 02138-1516, USA}

\author[0000-0001-5058-695X]{Jonathan P. Williams}
\affiliation{Institute for Astronomy, University of Hawai'i at M$\bar{\texttt{a}}$noa, Honolulu, HI, USA}

\begin{abstract}
We present evidence of variable 1.3~millimeter emission from the 1--3~Myr, SpT G2--G5 class~III YSO, HD~283572. 
HD~283572 was observed on 8 dates with the Submillimeter Array between 2021 December and 2023 May, a total on-source time of 10.2\,hours, probing a range of timescales down to 5.2 seconds.
Averaging all data obtained on 2022 Jan 17 shows a 4.4\,mJy ($8.8\sigma$) point source detection with a negative spectral index ($\alpha{=}{-2.7}{\pm}1.2$), with peak emission rising to 13.8\,mJy in one 3\,minute span, and 25\,mJy in one 29.7\,second integration ($L_\nu=4.7\times10^{17}$\,erg\,s$^{-1}$\,Hz$^{-1}$).
Combining our data for the other 7 dates shows no detection, with an rms noise of 0.24\,mJy\,beam$^{-1}$.
The stochastic millimeter enhancements on time frames of seconds--minutes--hours with negative spectral indices are most plausibly explained by synchrotron or gyro-synchrotron radiation from stellar activity.
HD~283572's 1.3\,mm light-curve has similarities with variable binaries, suggesting HD~283572's activity may have been triggered by interactions with an as-yet undetected companion.
We additionally identify variability of HD~283572 at 10~cm, from VLASS data.
This study highlights the challenges of interpreting faint mm emission from evolved YSOs that may host tenuous disks, and suggests that a more detailed temporal analysis of spatially unresolved data is generally warranted.
The variability of class~III stars may open up new ground for understanding the physics of flares in the context of terrestrial planet formation.

\end{abstract}



\section{Introduction}
\label{sec:intro}
Stellar flares are extreme radiation outbursts from the surfaces of stars, analogous to solar flares commonly observed on the Sun, whereby stored magnetic energy accelerates charged particles through surrounding stellar plasma, radiating brightly across the electromagnetic spectrum \citep{Dulk85, Fiegelson99, Gudel02, Fletcher11, Candelerasi14}.
At millimeter wavelengths, stellar variability campaigns have mostly targeted nearby main sequence M-type stars, such as Proxima Centauri and AU Mic, finding extreme flaring events on timescales of 1--10 seconds \citep{MacGregor20, MacGregor22, Howard23}. 
A handful of millimeter flares have also been detected from the closest Sun-like star, $\epsilon$ Eridani \citep{[$\epsilon$~Eri][]Burton22}, and a range of other M and K-type stars from millimeter survey telescopes \citep[e.g, the South Pole Telescope and the Atacama Cosmology Telescope, see e.g.,][]{Guns21, Naess21}.

\begin{table*}
    \centering
    \caption{SMA observational setup, over the 8 observing dates. The horizontal lines separate the survey in which HD~283572 was included as a target (2021--2022) from the dedicated follow-up campaign for variability (2023). All times represent time on source. The value of $\tau_{\rm{225}}$ represents the average opacity at 225\,GHz during observations.}
    \begin{tabular}{c|c|c|c|c|c|c|c|c|c}
         \hline
         \hline
         Date & Track & Scan time & N$_{\rm{scans}}$ & Flux & Bandpass & Phase & Antenna & N$_{\rm{ant}}$ & $\tau_{\rm{225}}$ \\
         & ID & [seconds] & & Calibrator(s) & Calibrator(s) & Calibrator(s) & Config. & &  \\
         \hline
         2021 Dec 24 & T1 & 29.7 & 120 & Uranus & 3c279 & 0510+180;\,3c111 & COM & 6 & 0.10 \\
         2022 Jan 14 & T2 & 14.8 & 222 & Ceres &3c279 & 0510+180;\,3c111 & COM & 6 & 0.05 \\
         2022 Jan 17 & T3 & 29.7 & 120 & Uranus &3c279 & 0510+180;\,3c111 & COM & 6 & 0.11 \\
         \hline
         2023 Mar 26 & T4 & 5.2 & 1080 & Uranus;\,Ceres & 1159+292 & 3c111 & SUB & 6 & 0.18 \\
         2023 Mar 27 & T5 & 5.2 & 1080 & Uranus;\,Ceres & 1159+292 & 3c111 & SUB & 6 & 0.15 \\
         2023 Mar 29 & T6 & 5.2 & 1080 & Ceres & 1159+292 & 3c111 & SUB & 6 & 0.22 \\
         2023 Mar 30 & T7 & 5.2 & 1080 & Ceres & 3c84;\,1159+292 & 3c111 & SUB & 6 & 0.20 \\
         2023 Mar 31 & T8 & 5.2 & 900 & Ceres & 3c84;\,1159+292 & 3c111 & SUB & 6 & 0.10 \\
    \hline
    \end{tabular}
    \label{tab:SMAobs}
\end{table*}


Stellar flaring rates decline with age due to stellar spin-down \citep{Davenport19}.
Decades of monitoring at radio and X-ray wavelengths have shown that young T-Tauri stars are highly active, and undergo intense periods of flaring \citep{Favata98, Stelzer07, Dzib15, Forbrich17, Vargas21}.
A small number of flares from T-Tauri stars have been confirmed at millimeter wavelengths \citep[e.g.,][]{Massi02,Bower03, Massi06, Massi08, Salter10, Mairs19}, but only recently are the statistics of millimeter variability coming to light \citep[e.g., from the ${\sim}$Myr-old Orion Nebular Cluster (ONC) and JCMT Gould-Belt transient surveys, see][respectively]{Vargas23, Johnstone+22}. 
Whilst millimeter flares have likewise been observed from a handful of classical RS~CVn binaries \citep[e.g.,][]{Beasley98, Brown06}, millimeter flares appear to be rarer than in X-ray or radio wavelengths, and also much stronger for young versus main sequence stars.
Prior millimeter studies have largely focused on `Classical T-Tauri Stars' (CTTSs, with significant disk accretion), whereas studies of Weak-Line T-Tauri Stars (WTTSs, with little-to-no active accretion) remain lacking.

HD~283572 is a G2--G5 SpT WTTS located in a 1--3\,Myr region (L1495) in Taurus (RA 04:21:58.85, Dec +28:18:06.51) at ${\sim}125.5$pc \citep{Blondel06, Wahhaj10, Gaia16, Gaia18, Gaia21, Krolikowski21, Anders22}. 
Accounting for the source distance, extinction ($A_{\rm{V}}=0.63$), metallicity ($[\rm{Fe/H}]=-0.2$), Gaia G--band magnitude ($8.8035{\pm}0.0029$), and B--R color ($1.068{\pm}0.006$), MIST model evolutionary tracks and isochrones \citep[with $0.1\,M_\odot$ mass spacing, and 0.1Myr age spacing;][]{Paxton11ApJS..192....3P, Paxton13, Paxton15, Dotter16, Choi16} demonstrate HD~283572 is likely $2.5{\pm}0.5$\,Myr with a mass $M_\star=1.4{\pm}0.1\,M_\odot$, i.e., a late-A/early-F type stellar precursor.
HD~283572 is a class~III young stellar object \citep[YSO; hosting little-to-no excess infrared emission][]{Lada84, Lada87, Adams87, WilliamsCieza11}, and is radio bright \citep[][]{ONeal90}. HD~283572 is also X-ray bright ($\log L_{\rm{X}}=29.8\,$erg\,s$^{-1}$ from 0.2--12\,keV, corrected to HD~283572's Gaia DR3 distance), and X-ray variable on timescales of 10s--100s of seconds, hosting extreme flaring events on $\approx$10,000\,s timescales \citep{Favata98, Scelsi05, Pye2015}.

Here we report an extreme millimeter brightening event associated with HD~283572 detected with the Submillimeter Array (SMA) found serendipitously during a survey for class~III circumstellar dust disks.
We interpret this as millimeter stellar activity, amongst the first reported for a class~III YSO. 
We present the SMA observations and analysis in \S\ref{sec:obs}, our discussion in \S\ref{sec:discussion}, and conclusions in \S\ref{sec:conclusions}.

\section{SMA observations and analysis} \label{sec:obs}
\subsection{SMA data calibration}
We observed HD~283572 with the SMA \citep[][]{Ho04} on the dates listed in Table~\ref{tab:SMAobs} between 2021--2023, over ``tracks'' of 2--10\,hours with 6 operable antennas. 
For all tracks, the receivers were tuned to an LO frequency of 225.1 or 225.5\,GHz ($\lambda=1.33$\,mm); providing frequency coverage from 209.1--221.1\,GHz (or 209.5--221.5\,GHz; lower sideband) and 229.1--241.1\,GHz (or 229.5--241.5\,GHz; upper sideband).
The SWARM correlator processed the 48\,GHz available bandwidth, all with channel spacing of 140\,kHz.
Standard calibrator sources and setups were used with each track. 
Observations were conducted in two configurations (COM and SUB, with angular resolutions of 2--3$''$ and 4--5$''$ respectively). Two distinct modes of time sampling were employed. The first mode used integrations of 14.8 or 29.7\,secs over 2--3\,mins, 2--3 times per hour over 9--10\,hr time spans (cycling between different sources and phase calibrators). The second mode (DDT follow-up) used shorter integrations of 5.2\,secs over ${\sim}$20\,mins between phase calibrator observations, for 2--3\,hrs.

\begin{figure*}
    \centering
    \includegraphics[width=1.0\linewidth]{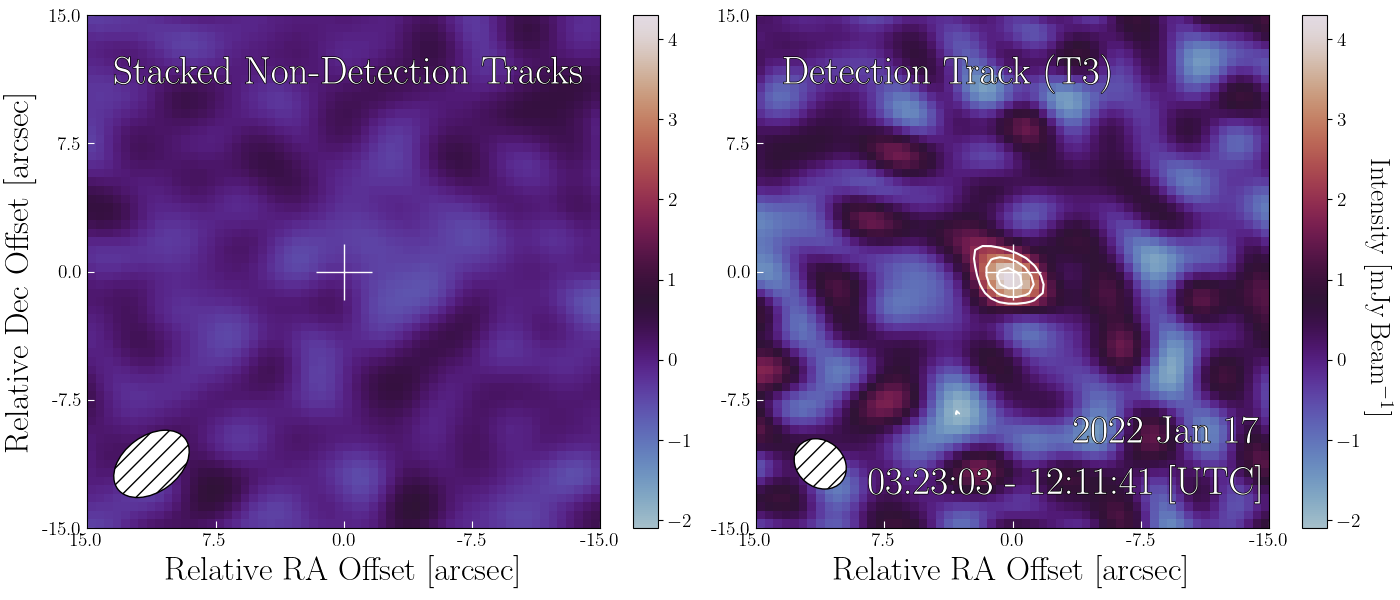}
    \caption{\textit{Left}: cleaned image of the combined SMA measurement sets from all tracks in which no significant emission is detected. 
    \textit{Right}: cleaned image of the SMA measurement set for track T3, in which a significant point source is present at the location of HD~283572.
    The noise is lower in the image with no detection ($\rm{RMS}_{\rm{combined}}$=0.24\,mJy\,beam$^{-1}$ than the image with the detection ($\rm{RMS}_{\rm{T3}}$=0.48\,mJy\,beam$^{-1}$). 
    Contours are at ${\pm}4$, 6 and 8$\sigma$ levels. SMA beams are in the lower-left.}
    \label{fig:trackVariability}
\end{figure*}

All data (see \S\ref{sec:data}) was re-binned by a factor of 32 to reduce data sizes and speed up calibration (appropriate for continuum analysis).
We converted the raw data to the \textit{Common Astronomy Software Applications} \citep[{\tt CASA};][]{CASA} measurement set format with the {\tt PYUVDATA} \citep{pyuvdata} SMA reduction pipeline\footnote{Both SMA {\tt CASA} and {\tt MIR} reduction pipelines can be accessed here: https://github.com/Smithsonian/sma-data-reduction} in {\tt CASA} pipeline v6.4.1.
We manually flagged the limited number of narrow interference spurs, as well as the outermost 2.5\% ``guard-band'' channels from each spectral window.
After calibration, the `corrected' data were spectrally averaged to 4 channels per window to expedite our analysis with $\tt{mstransform}$. All images were produced using natural weighting (to maximize signal-to-noise) with CASA's $\tt{tclean}$ task.

\subsection{SMA data analysis}
\label{sec:smaVarAnalysis}
\subsubsection{Per-track variability}
We image each of the 8 tracks separately: observations of HD~283572 in tracks T1, T2, and T4--T8 show no significant emission (see Fig.\ref{fig:trackVariability}, left), neither in their imaged data nor in visibility space. Combining data from all 7 tracks similarly show no significant emission, with an estimated RMS noise of $240\,\mu$Jy\,beam$^{-1}$, and no peaks exceeding ${\pm}4\sigma$ anywhere.

For T3, however, significant emission was detected in both the image and visibility fit (see Fig.\ref{fig:trackVariability}, right). 
Imaging analysis (via CASA's {\tt imfit}) returns a peak emission of $4.2{\pm}0.5$\,mJy\,beam$^{-1}$ centered on HD~283572's location. Visibility fitting (via CASA task {\tt uvmodelfit} with a point source model) returns  $4.4{\pm}0.5$\,mJy, centered at $\Delta \rm{RA}=0.42''$, $\Delta \rm{Dec}=-0.67''$, consistent with the position of HD~283572 given the astrometric uncertainty of the SMA data.
The significance of the detection in the single track T3 image is $8.2\sigma$ (and $8.8\sigma$ for the visibility-based fit). Fitting for the upper and lower sidebands (209.1--221.1\,GHz and 229.1--241.4\,GHz, respectively) of the SMA data independently, we determine a spectral index of $\alpha_{\rm{T3}}=-2.7{\pm}1.2$, where $\alpha := \partial \log{F_\nu}/ \partial \log{\nu}$.

The non-detection in T2 and T4, and strong detection in T3 implies that the stellar activity rose significantly in the three days between 2022 Jan 14 and Jan 17, and fell on an uncertain timescale by 2023 Mar 26.
There is no evidence of periodicity on the time frame of days given the non-detections T4--T8, nor on weeks--months timescales given the sparse coverage in observations.


\begin{figure*}
    \centering
    \includegraphics[width=1.0\linewidth]{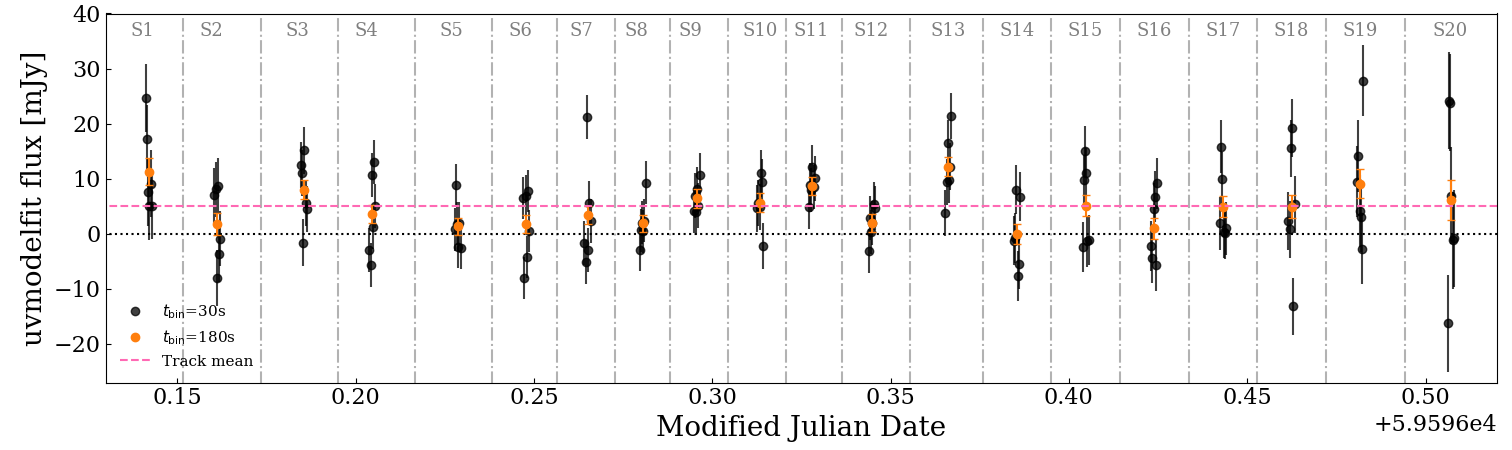}
    \caption{Flux versus time for Track T3, at 29.7\,s (black) and 180\,s (orange) resolution, including the track-averaged flux (pink, dashed).
    Shown are the 20 regions we refer to as segments from S1 to S20. All other tracks show data consistent with noise, whereas this track shows persistent elevated emission, with a significant brightening event during S13.}
    \label{fig:lightcurve}
\end{figure*}

\subsubsection{Per-scan variability}
We further examine the millimeter variability \textit{within} each track.
To aid our initial comparison, we fit point source fluxes (with fixed offsets $\Delta \rm{RA}=0.4''$ and $\Delta \rm{Dec}=-0.7''$, consistent with the best-fit from analysis of T3) to the visibilities in all 8 tracks, binned at either 30s, 60s, and 180s resolution, an approach consistent with the studies of \citet{MacGregor20} and \citet{Burton22}. 
These represent time-binning by factors of 1--6 (for T1 and T3); 2--12 (for T2); and 6--36 (for T4--T8).
Since this represents significant re-binning for tracks T4--T8 which were observed at much higher temporal resolution, we also fit fluxes for these 5 tracks at 5.2s (full) and 10.4s temporal resolution, finding no significant emission peaks.
Whilst fixing the offset may reduce the significance of individual brightening events since the position is consistent with HD~283572's location (and the peak of emission detected in T3), we expect this to still robustly detect significant brightening events.
We show in Fig.~\ref{fig:lightcurve} the resultant lightcurve for T3.
By comparing these fitted flux values with their uncertainties on timescales of 30s--180s, we found one significant brightening event, evident throughout `segment 13' (S13) of T3, with (sub-)segments of other tracks being consistent with noise.

\begin{figure*}
    \centering
    \includegraphics[width=1.0\linewidth]{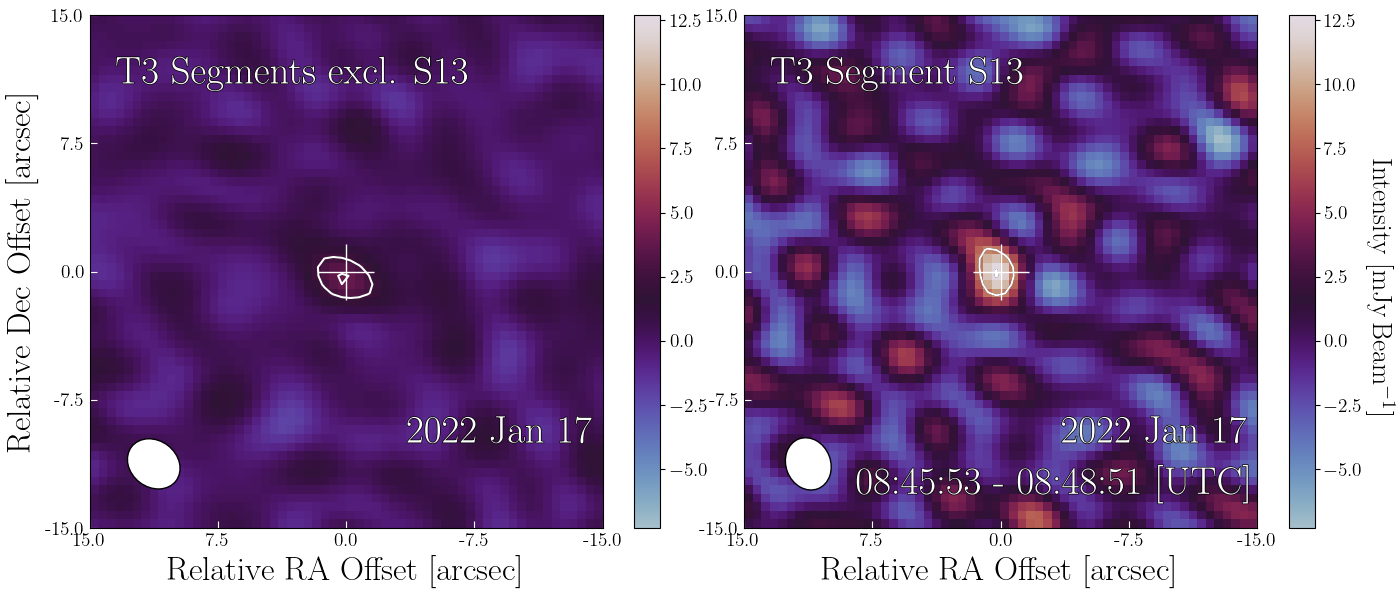}
    \caption{\textit{Left}: cleaned image of T3 with T3:S13 excluded in which a bright point source is present at HD~283572's location. 
    \textit{Right}: cleaned image of T3:S13 on its own, in which an even brighter point source is present at HD~283572's location. 
    The rms in the left and right images are 0.64\,mJy\,beam$^{-1}$ and 2.1\,mJy\,beam$^{-1}$ respectively, and so both are significant detections, albeit at different flux levels. Contours are at ${\pm}4$ and 6$\sigma$ levels. SMA beams are in the lower-left.}
    \label{fig:T3Variability}
\end{figure*}

T3 segment 13 (T3:S13) spanned the time interval 2022 Jan 17, 08:45:52.9--08:48:51.0 UTC.
Imaging T3:S13 using the {\tt tclean} task (to $2\,\sigma$ threshold with a central $5''$ radial mask), we find (using {\tt imfit}) that the total (unresolved) flux co-located with HD~283572 is $13.8{\pm}1.8$\,mJy, a detection of the brightening event at the $7.7\sigma$ level.
Imaging track T3 excluding the T3:S13 segment, we find (using {\tt imfit}) an unresolved flux of $3.8{\pm}0.4$\,mJy.
Both images are presented in Fig.~\ref{fig:T3Variability}.
Fitting again for the upper and lower sidebands of the data separately, we find the emission during T3:S13 to have a negative spectral index of $\alpha_{\rm{F1}}=-3.0{\pm}2.0$ and $\alpha_{\rm{T3\,minus\,F1}}=-1.0{\pm}1.2$ in the full track T3 excluding T3:S13 data, consistent with this remaining fixed throughout the entire track.
Although cross-hand polarization data were not collected (XY and YX), we are able to leverage the changing parallactic angle of the source to set a constraint on the total linear polarization fraction ($p_{Q{+}U}$) during the T3 track of $p_{Q{+}U}=0.21\pm0.21$, consistent with a null detection.
We measured the T3:S13 fluxes in the XX and YY polarization channels as $F_{\rm XX, T3:S13}=12.6{\pm}2.3\,$mJy and $F_{\rm YY, T3:S13}=18.7{\pm}2.5\,$mJy, also consistent with a null detection.

\begin{figure*}
    \centering
    \includegraphics[width=1.0\linewidth]{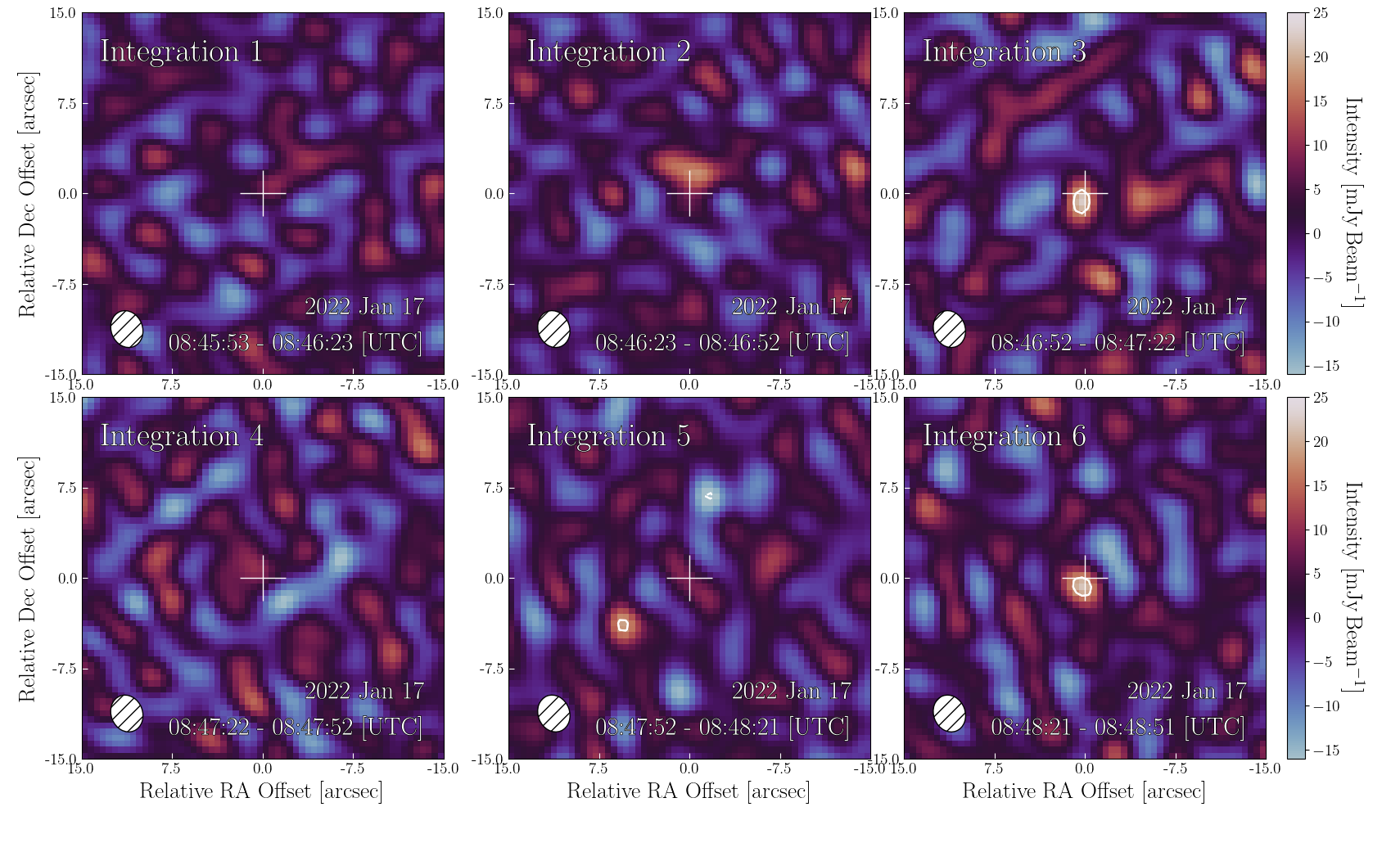}
    \caption{From left--right, then top--bottom: maps of the 29.7 integrations during Track T3, segment 13 (T3:S13; when the event F1 was detected). Image rms values range from 3.8--4.8\,mJy\,beam$^{-1}$. 
    Integrations 3 and 6 show emission at the 23mJy and 25mJy level respectively coincident with HD~283572's location. 
    Contours are at ${\pm}4\sigma$ levels. SMA beams are in the lower-left.}
    \label{fig:flaresubscans}
\end{figure*}

In Fig.\ref{fig:flaresubscans} we present images of all six scans during T3:S13, imaged at their native 29.7\,sec resolution. As can be seen, there is emission present at the location of HD~283572 in integrations 3 and 6.
Applying the same $\tt{imfit}$ routine to these images, we measure brightness peaks of $25.0{\pm}3.8$\,mJy and $23.0{\pm}4.8$\,mJy respectively (and later refer to these two events as `F1A' and `F1B' respectively). This method only yielded 2--4$\sigma$ significant fits for the T3:S13 integrations 1, 2, 4 and 5 separately.
Whilst these results may suggest that the majority of the observed flux over the 3\,min cadence T3:S13 arose from just two 29.7\,sec integrations, our data are unable to determine the underlying variability during F1.
Further, whilst integrations 1, 2, 4 and 5 have an average flux $11.5{\pm}2.3\,$mJy, and spectral index $\alpha=-0.3{\pm}3.2$, the spectral indices during events F1A and F1B are $\alpha_{\rm{F1A}}=-11{\pm}5$ and $\alpha_{\rm{F1B}}=-0.5{\pm}5.0$ respectively, both negative albeit with large uncertainties and our data are thus consistent with having a fixed spectral slope during event F1.

\subsubsection{Summary of analysis}
We summarize four significant findings.
First, during an active-period (T3) HD~283572's emission was significantly elevated for at least 9\,hours (which we refer to as HD~283572's `active-quiescent' level).
Second, during this active-period in the 3-minute segment T3:S13, HD~283572's emission substantially increased, (which we refer to as `F1'). The difference in flux of `F1' and the active-quiescent level ($10.0{\pm}1.8$\,mJy) is statistically significant ($5.5\sigma$), and represents a flux enhancement by a factor of around 3.6.
Third, in HD~283572's active-quiescent period, its flux rose by ${\gtrsim} 16 \times$ over the inactive period, and by ${\gtrsim} 60\times$ during F1. 
Fourth, the millimeter spectral index is consistently steep and negative during F1 and the active-quiescent period.
We tabulate all observed and derived flare properties in Appendix~\ref{sec:flareproperties}, Table~\ref{tab:flareproperties}.

\section{Discussion}
\label{sec:discussion}
In \S\ref{sec31} we first discuss the derived millimeter/radio properties of HD~283572 implied by our observations, compare these with the literature of other millimeter flares in \S\ref{sec32_2}, and explore future implications in \S\ref{sec:nextsteps}.
\subsection{HD~283572: a powerful millimeter flaring star} \label{sec31}
\subsubsection{Stellar activity: timescales and strengths}
Millimeter circumstellar dust emission around class~III stars is typically faint (${<}1\,$mJy at the distances of nearby ($\sim140$\,pc) star-forming regions), has a positive spectral index and is stable in flux on timescales spanning years \citep[e.g.,][]{Wyatt08, Matthews14,  Hughes18, Lovell21a}, whereas emission from stellar flares can show enhancements on timescales of seconds to hours, with steep negative spectral indices. The stochastic millimeter emission of HD~283572 is most plausibly explained as resulting from stellar activity. 

In assessing the flare frequency rate, we note that the SMA observed HD~283572 for a total on-source time of 10.2\,hours, within which we detected one (sparsely sampled) active-quiescent period and one F1-type event. 
From this we estimate the rate of active periods as $0.001-0.9$\,day$^{-1}$, during which the rate of F1-type flare events is $0.005-0.1$\,hour$^{-1}$, accounting for the fact that the sparse sampling may have missed the beginning and ending of these events (for the active-quiescent period, the event may extend anytime between the end of T2 to the start of T4; 463\,days, and for F1, from the end of T3:S12 to the start of T3:S14; 58\,mins).

We determine a range of millimeter luminosity densities, $L_{\rm{mm}} = 4\pi d^2 F_{\rm{\nu=225GHz}}$.
For the active-quiescent period, $L_{\rm{mm}} = 7.2\times10^{16}$\, erg\,s$^{-1}$\,Hz$^{-1}$, for event F1, $L_{\rm{mm}} = 2.6\times10^{17}$\, erg\,s$^{-1}$\,Hz$^{-1}$, and for events F1A and F1B, $L_{\rm{mm}} = 4.7\times10^{17}$ erg\,s$^{-1}$\,Hz$^{-1}$ and $L_{\rm{mm}} = 4.3\times10^{17}$ erg\,s$^{-1}$\,Hz$^{-1}$ respectively.
We place an upper-limit on the millimeter luminosity density for the inactive period of $L_{\rm{mm}} {<} 4.5\times10^{15}$\, erg\,s$^{-1}$\,Hz$^{-1}$ (a factor of ${\sim}100$ below the F1A peak), and also present these in in Appendix~\ref{sec:flareproperties}, Table~\ref{tab:flareproperties}.

Data from the Karl G. Jansky Very Large Array (VLA) Sky Survey \citep[VLASS;][]{Lacy2020} shows radio emission associated with HD~283572 in 2019 and 2021 (which we present in detail in Appendix~\ref{sec:appendixVLASS} to investigate HD~283572's radio emission properties, outside of our SMA-observation window).
We determine the 3\,GHz radio luminosity density of HD~283572 from VLASS epoch 1.2 and 2.2 as $L_R = 4\times10^{16}$\,erg\,s$^{-1}$\,Hz$^{-1}$ and $L_R = 9\times10^{15}$\,erg\,s$^{-1}$\,Hz$^{-1}$ respectively. We find HD~283572's X-ray luminosity \citep[$\log L_{\rm{X}}{\sim}29.4$\,erg\,s$^{-1}$;][]{Favata98,Pye2015} to be a factor of 3--12 above the G\"udel-Benz relationship \citep[well within the scatter of][i.e., $L_{\rm{X}}/L_{\rm{R}} = \kappa\times10^{15.5{\pm}0.5}$]{Benz93, Benz94}, assuming $\kappa=0.03$, as shown by e.g., \citet{Dzib13, Dzib15} to fit populations of YSOs in Ophiuchus and Taurus.
Consistency with the G\"udel-Benz relationship suggests that HD~283572 is a magnetically active young star.

\subsubsection{Emission sources} \label{sec32}
The millimeter spectral indices of T3, F1, and T3 excluding F1 are all negative, with $\alpha_{\rm{T3}}{=}-2.7{\pm}1.2$, $\alpha_{\rm{F1}}{=}-3.0{\pm}2.0$ and $\alpha_{\rm{T3\,minus\,F1}}{=}-1.0{\pm}1.2$.
By extrapolating the flux implied by these slopes \citep[and accounting for the typical ${\sim}1{-}10\,$GHz turnover in the gyro-synchrotron spectrum, see e.g.,][]{Gudel02} during a flare HD~283572's 3\,GHz emission should plausibly reach 10s--100s of mJy. 
We do not observe fluxes this high in the VLASS data, suggesting HD~283572 was in an inactive period during these measurements. 
We can constrain the dominant emission source probed by the VLA with our SMA measurements by measuring the millimeter-radio spectral index, anchored by the SMA upper-limit of 0.24\,mJy.
We estimate $\alpha$ as ${<}{-0.2}$, ${<}{-0.5}$ and ${<}{-0.7}$ respectively for VLASS 2.2, 1.2 and a 5\,GHz 1989 observation of HD~283572 \citep{ONeal90} by assuming a power-law flux scaling of $F_\nu \propto \nu^\alpha$ for the three VLA measured 0.5, 2.1\,mJy (3GHz) and 3.3\,mJy (5GHz) fluxes.
These spectral indices are consistent with non-thermal gyro-synchrotron/synchrotron emission \citep{Gudel02}, which at millimeter wavelengths is likely optically thin \citep{Klein87}.
This emission originates from electrons gyrating along stellar magnetic field lines after magnetic reconnection events in stellar magnetospheres and energize non-thermal particles on rapid timescales.

We associate a brightness temperature with the SMA emission via $T_b \approx \frac{1}{2k_b}(\frac{d}{r})^2 \lambda^2 \Delta S$ \citep[see e.g.,][equation 3, for distance $d$, emission radius $r$, wavelength $\lambda$, and flux enhancement $\Delta S$]{Gudel02}, by assuming emission emanates from length scales ranging from a few Earth radii (comparable to Solar active regions) to a few Solar radii (comparable to flaring loops).
We find that event F1 is associated with a brightness temperature in the range $T_b\approx 10^8 -10^{10}$\,K, in accordance with incoherent gyro-synchrotron/synchrotron emission. 

Synchrotron and gyro-synchrotron emission have different millimeter properties, however we are unable to distinguish between these with our observations.
The optically thin spectral slopes of synchrotron and gyro-synchrotron emission are defined by $\alpha{=}1.22{-}0.9\delta_{GS}$ and $\alpha{=}{-}(\delta_S{-}1)/2$, with gyro-synchrotron emission being moderately circularly polarized, and synchrotron emission linearly polarized with fraction $r{=}(\delta_S{+}1)/(\delta_S{+}(7/3))$. 
From T3, our measured $\alpha{=}{-}2.7{\pm}1.2$ thus translates to either $\delta_S{=}6.4$ or $\delta_{GS}{=}4.4$ for synchrotron/gyro-synchrotron respectively. 
Our null detection of linear polarization (with fraction $p_{Q{+}U}{=}0.21\pm0.21$) is thus broadly inconsistent with the expected fraction ($r(\delta_S){=}0.85$) if the emission was synchrotron, suggesting gyro-synchrotron may be favored. 

The strength of the magnetic field that we estimate as needed for gyro-synchrotron emission to radiate at 225\,GHz however seems too high.
We estimate this by associating the SMA observation frequency with a harmonic ($s$) of the gyro-frequency of emission, i.e., $\nu_0 = 2.8\times10^6 B \gamma^2$\,Hz \citep[for the peak spectral frequency $\nu_0$;][]{Gudel02}.
For gyro-synchrotron, $s{=}10{-}100$ and $\gamma{\lesssim}2{-}3$, whereas for synchrotron, $s{>}100$ and $\gamma{\gg}1$.
For example, in a gyro-synchrotron case ($\gamma{=}3$) we find $B{\gtrsim}9\,$kG, whereas in a synchrotron case (${\gamma}{=}10$) we find $B{\lesssim}1\,$kG. 
The lower value appears more consistent with expectations from low--intermediate mass YSO B-fields \citep[see e.g.,][]{Folsom16}, and thus may instead imply that these HD~283572 SMA observations probed a synchrotron emission mechanism.

\subsection{Comparison with published millimeter flares} \label{sec32_2}
There are few published detections of millimeter flares that constrain their luminosities, timescales, spectral slopes, and polarization. Here we briefly compare/contrast HD~283572's stellar activity with available published constraints.

\subsubsection{Flares from single stars}
The details of HD~283572's millimeter emission appear different than those reported for main sequence flares from single stars, such as those of AU~Mic, Proxima~Centauri, $\epsilon$~Eri and the Sun \citep[][ with millimeter luminosity densities of $10^{13}$--$3\times10^{14}$\,erg\,s$^{-1}$\,Hz$^{-1}$, 10s of seconds flare timescales, and for the Sun and $\epsilon$~Eri, positive/flat spectral indices]{Krucker13, Macgregor18, MacGregor20, Burton22}.
HD~283572's flare luminosity density is a factor of ${\sim}1000\times$ higher than the brightest of those, $\epsilon$~Eri (based on F1/F1A), persisted over ${\gtrsim}9\,$hours (${\sim}1000\times$ longer) and had a negative spectral index throughout.

The Orion-based class~III YSO, GMR-A showed a powerful stellar flare \citep[][]{Bower03}, with properties more closely matching those of HD~283572. 
Nevertheless, the power ($\nu L_\nu$) associated with GMR-A was around 10$\times$ higher than what we report for HD~283572, and its spectral slope was found to change over time (whereas we find no evidence of variability in HD~283572's spectral slope).

\subsubsection{Flares from binary/multiple stellar systems}
A number of YSO millimeter/radio flares have been inferred to result from periodic magnetospheric interactions within stellar multiples \citep[e.g., V773~Tau, DQ~Tau and JW~566, see][respectively, all class~II systems, with the flaring periodicity of V773~Tau and DQ~Tau being directly tied to known orbital periastron-passages]{Massi06, Salter10, Mairs19}.
In each case, their peak luminosity densities reached comparable/higher levels than HD~283572 (by a factor of a few to $\approx100$), their event times spanned hours (although JW~566 is limited by the JCMT cadence of several days), however their spectral indices were either positive (DQ~Tau), flat (JW~566) or unreported \citep[V773~Tau, although][infer this to be synchrotron emission which would have a negative spectral index]{Massi06}.

The RS Canum Venaticorum (RS~CVn) variable binary star, $\sigma$~Gem showed a millimeter flaring event in 2004 \citep[][]{Brown06}.
$\sigma$~Gem presented non-detections over many days preceding its flare, became active over a period of several hours (rising to a peak lasting only minutes, with a power lower than that of HD~283572's by a factor of 7), returned to a detected active-state, and then in subsequent observations was undetected. This light-curve in particular has a temporal profile similar to HD~283572.
Since the $\sigma$~Gem measurements lack polarization or spectral index constraints, we are unable to compare their properties further.

\subsubsection{Detections from cosmology telescopes}
Bright stellar flares have been measured during deep millimeter observations with wide-field cosmology telescopes \citep[e.g., the South Pole Telescope, SPT, and the Atacama Cosmology Telescope, ACT, see][respectively]{Guns21,Naess21}.
These surveys detected 90/150/220\,GHz positive/flat spectral index sources with $L_\nu \approx 10^{16}-10^{19}$\,erg\,s$^{-1}$\,Hz$^{-1}$ that are associated with stars, over timescales spanning minutes--days. 
These luminosity densities are consistent with HD~283572's, however the positive/flat spectral indices are inconsistent.
The SPT/ACT cadences make it difficult to compare with HD~283572's light-curve. 

\subsubsection{Is HD~283572 a binary?}
Overall, HD~283572's millimeter activity appears to more closely match systems with companions, despite HD~283572 having no evidence of binarity \citep{Krolikowski21}.
We however cannot rule out the possibility that HD283572 is in an unresolved binary system with interactions with such an unresolved companion having triggered flares/stellar activity.
HD~283572's Gaia RUWE (0.925) implies that any binary companion would need to be low-mass and on a short-period orbit, which could easily have gone undetected given the star's large rotational broadening \citep[i.e., $v \sin{i}=110{\pm}20$\,km\,s$^{-1}$, see e.g.,][]{Fernandes98}. 
HD~283572's mass of ${\approx} 1.4\,M_\odot$ (estimated in \S~\ref{sec:intro}) suggests it will become an early-F/late-A-type main sequence star, which statistically favors binarity.
New observations are needed to ascertain the multiplicity of HD~283572, and then further whether a binary interaction triggers HD~283572's flares. 

\subsection{Class~III stars: ideal probes of stellar flare physics}
\label{sec:nextsteps}
There is a rich history of centimeter-wavelength radio studies of WTTSs (which almost fully overlap with the population of class~III YSOs).
VLA observations have shown these stars can be radio-bright \citep[e.g.,][]{ONeal90, White92} and more recently, enhanced in both their brightness and variability versus less-evolved YSOs \citep[see e.g. Figures 3 and 4 of][]{Dzib13, Dzib15}.
The increase in class~III YSO radio variability make them ideal for systematic searches for millimeter-counterpart flares \citep[e.g., utilizing new \textit{Gaia} population analyses, see][for which class~III stars typically dominate YSO populations in star-forming regions]{Luhman20,Krolikowski21,Luhman22a,Luhman22b,Luhman23a,Luhman23b}.

Our analysis suggests that temporal analyses of class~III YSO millimeter observations are generally warranted to distinguish between stellar and circumstellar components.
Importantly, class~III stars open a window on planet formation/disk evolution after the dispersal of the \textit{bulk} protoplanetary/primordial disk material. 
Class~III YSOs can retain large reservoirs of cold dust, and/or smaller reservoirs of hot dust that trace ongoing terrestrial planet formation processes \citep[][]{Lovell21a,Michel21} over timescales that fully span pre-main sequence evolution \citep[][]{Kenyon04b, Morbidelli12}. Combined with the implication that proton-rich flare events can disrupt the growth of planetary atmospheres \citep{Tilley19}, constraints on the physics of class~III YSO flares could provide key inputs to exoplanet atmospheric growth models.

\section{Conclusions} \label{sec:conclusions}
We present new observations of HD~283572 with the Submillimeter Array (SMA) taken from 2021--2023, alongside VLA data taken from 2019--2021.
We show that HD~283572 is variable at millimeter and radio wavelengths, and underwent one extreme millimeter brightening event on 2022 Jan 17.
Our analysis suggests HD~283572's millimeter variability was due to an extreme stellar flare, during a prolonged active-quiescent period that was both preceded and followed by inactive periods.
By constraining HD~283572's spectral index, brightness temperature, and linear polarization fraction, we find that the emission is most likely from gyro-synchrotron/synchrotron radiation. HD~283572's stellar activity appears broadly similar to that reported for variable binaries based on their peak luminosities, timescales/light-curves and spectral slopes. These results suggest that HD~283572's activity may have been induced by interactions with an as-yet undetected companion. 
Although currently a class~III YSO, HD~283572 will likely evolve into a late-A/early F-type star on the main sequence, indicating that intermediate mass stars can be active at millimeter wavelengths at young ages.
Millimeter observations of class~III YSOs/WTTTs provide an opportunity to study the physics of stellar flares, and their implications for terrestrial planet formation/planetary atmospheric growth.

\begin{table*}
    \centering
    \caption{Flare properties based on all observations of HD~283572, including the average flux over the observation timescale, associated luminosity density (at the distance to HD~283572) $L_\nu$, spectral index $\alpha$, and (where this can be measured) the linear polarization fraction $p_{Q+U}$.}
    \begin{tabular}{l|c|c|c|c|c}
         \hline
         \hline
         ID & Flux & Timescale & $L_{\nu=225\,\rm{GHz}}$ & $\alpha$ & $p_{Q+U}$ \\ 
          & [mJy] & [UTC] & [erg\,s$^{-1}$\,Hz$^{-1}$] & & \\
        \hline
          T3 & $4.4{\pm}0.5$ & 03:23:03.0--12:11:41.0 (8.81\,hours) & $8.3\times10^{16}$ & $-2.7{\pm}1.2$ & $0.21{\pm}0.21$ \\ 
        T3:S13 (F1) & $13.8{\pm}1.8$ & 08:45:52.9--08:49:51.0 (3\,mins) & $2.6\times10^{17}$ & $-3.0{\pm}2.0$ & --\\
        T3:S1--12 and T3:S14--20 & $3.8{\pm}0.4$ & (8.76\,hours, excl: F1) & $7.2\times10^{16}$ & $-1.0{\pm}1.2$ & $0.09{\pm}0.22$ \\
        T3:S13:I3 (F1A) & $25.0{\pm}3.8$ & 08:46:52.3--08:47:22.0 (29.7\,secs) & $4.7\times10^{17}$ & $-11{\pm}5$ & --\\
        T3:S13:I6 (F1B) & $23.0{\pm}4.8$ & 08:49:21.3--08:49:51.0 (29.7\,secs) & $4.3\times10^{17}$ & $-0.5{\pm}5.0$ & --\\
        T3:S13:I1--2 and T3:S13:I4--5 & $11.5{\pm}2.3$ & (2\,mins, excl: F1A and F1B) & $2.2\times10^{17}$ & $-0.3{\pm}3.2$ & --\\
        \hline
          T1,T2,T4,T5,T6,T7 and T8 & ${<}0.72$ ($3\sigma$) & \textit{Multiple (see Table~\ref{tab:SMAobs})} & ${<} 4.5\times10^{15}$ & -- & -- \\
        \hline
    \end{tabular}
    \label{tab:flareproperties}
\end{table*}

\section{Software and third party data repository citations} \label{sec:data}
The SMA data used here are from projects 2021B-H003, 2021B-S014 and 2022B-S069 and can be accessed via the Radio Telescope Data Center (RTDC) at \url{https://lweb.cfa.harvard.edu/cgi-bin/sma/smaarch.pl} after these have elapsed their proprietary access periods (note only data that passed strict quality control checks was included in this work).
The VLA data used here are from the VLASS project, epochs `VLASS1.2' and `VLASS2.2', and collected from the Canadian Astronomy Data Centre (CADC).
The National Radio Astronomy Observatory (NRAO) is a facility of the National Science Foundation (NSF) operated under cooperative agreement by Associated Universities, Inc. 
IRADA is funded by a grant from the Canada Foundation for Innovation 2017 Innovation Fund (Project 35999), as well as by the Provinces of Ontario, British Columbia, Alberta, Manitoba and Quebec.
This work has made use of data from the European Space Agency (ESA) mission Gaia (\url{https://www.cosmos.esa.int/gaia}), processed by the Gaia Data Processing and Analysis Consortium (DPAC, \url{https://www.cosmos.esa.int/web/gaia/dpac/consortium}).
Funding for the DPAC has been provided by national institutions, in particular the institutions participating in the Gaia Multilateral Agreement.

\vspace{0mm}
\facilities{Submillimeter Array (SMA), Karl G. Jansky Very Large Array (VLA)}

\software{This research made use of a range of software packages, highlighted here:
astropy \citep{2013A&A...558A..33A,2018AJ....156..123A},
          CASA \citep{CASA},
          pyuvdata \citep{pyuvdata}.
}

\vspace{-3mm}
\begin{acknowledgments}
J. B. Lovell acknowledges the Smithsonian Institute for funding via a Submillimeter Array (SMA) Fellowship, and (although not ultimately presented in this work) thanks both Lizhou Sha and Alexander Binks for discussions on TESS data and its analysis. 
The authors thank the anonymous referee for useful comments which substantially improved the content and clarity of this work.
The authors wish to recognize and acknowledge the very significant cultural role and reverence that the summit of Maunakea has always had within the indigenous Hawaiian community, where the Submillimeter Array (SMA) is located. 
We are most fortunate to have the opportunity to conduct observations from this mountain.
We further acknowledge the operational staff and scientists involved in the collection of data presented here.
\end{acknowledgments}

\appendix{}
\section{Derived flare properties}
\label{sec:flareproperties}
We collate in Table~\ref{tab:flareproperties} the observed and derived properties of HD~283572 during our SMA observations, including the ID associated with the emission, the average flux, time/timescale of the event/s, luminosity density, spectral index and the linear polarization fraction (in the case of the full track, and full track minus event F1, where sufficient parallactic angle coverage (during the flare) is available to unambiguously constrain this fraction). We present all values here since some derived values are not strictly independent (e.g., in the case of T3 includes data from T3:S13), whereas `T3:S13 (F1)' is fully independent of the data in `T3:S1--12 and T3:S14--20'.

\section{VLASS counterpart radio variability}
\label{sec:appendixVLASS}
\begin{figure}
    \centering
    \includegraphics[width=1.0\linewidth]{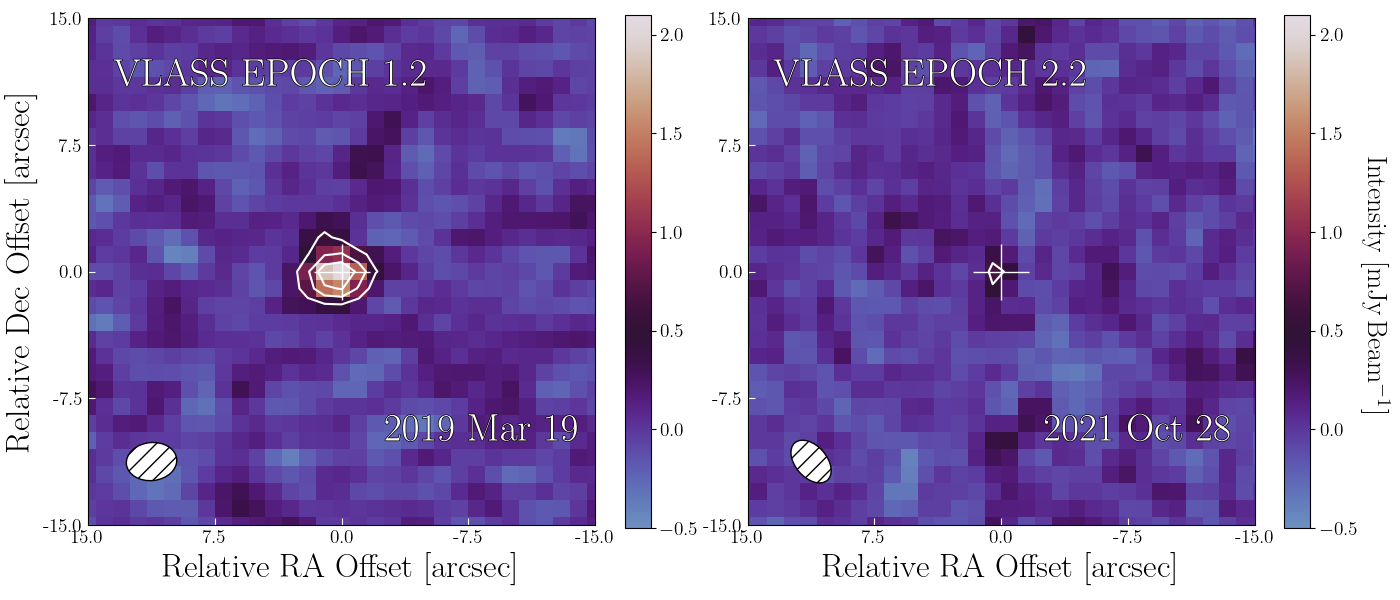}
    \caption{VLASS `Quick Look' images (left: epoch 1.2; right: epoch 2.2).In both images, the coordinates are centred on the location of HD~283572. Clear in 1.2 is the detection of a bright point source, whereas in 2.2 the significance is much reduced. The rms in both images is 0.125\,mJy\,beam$^{-1}$. Contours are at ${\pm}4$, 8 and 12 $\sigma$ levels. VLA beams are in the lower-left.}
    \label{fig:VLASSfits}
\end{figure}

The Very Large Array (VLA) has observed HD~283572 on a number of occasions, with \citet{ONeal90} observing first at 5\,GHz, measuring a flux of $3.29{\pm}0.28$\,mJy.
Recently, the VLA re-observed HD~283572 during the VLA Sky Survey \citep[VLASS;][in two epochs at 2--4\,GHz, $2.5''$ resolution, to a depth of ${\sim}120\,\mu$Jy]{Lacy2020}.
Using the multi-epoch version of $\tt{SODA}$ ($\tt{SODA}$ is designed to extract single-epoch cutouts of `quicklook' VLASS images, as produced by Mark Lacy, for which the code is provided here: \url{https://gitlab.nrao.edu/mlacy/vlass-vo/-/blob/main/SODA_multi_pos_multi_ep.ipynb}) we sourced fits images centred on the coordinates of HD~283572 (in epochs `VLASS1.2' and `VLASS2.2', observed on dates of 2019 Mar 19 and 2021 Oct 28 respectively), which both showed unresolved emission in the central pixels coincident with HD~283572's location, which we present in Fig.\ref{fig:VLASSfits}.
We fitted 2D Gaussian ellipses (within $5''$ of HD~283572's location) using {\tt ASTROPY GAUSSIAN2D}, and found HD~283572 to host unresolved integrated fluxes of $2.13{\pm}0.13$\,mJy and $0.49{\pm}0.13$\,mJy (epochs 1.2 and 2.2 respectively).
Our analysis demonstrates HD~283572 is highly variable at radio wavelengths on 2-year timescales, being dimmer by $4.3\times$ in 2021 (the immediate run-up to our SMA observations) versus 2019.

\bibliography{sample631}{}
\bibliographystyle{aasjournal}
\end{document}